\documentclass{article}

\input{epsf}

\begin{document}

\fontsize{12}{14}\selectfont

\centerline{\Large {\bf  Dynamical partitions of space in any dimension}}
\vskip.7cm

\centerline{{Tomaso Aste } }

\vskip.7cm
\centerline{\it AAS, Sal. Spianata Castelletto 16, 16124 Genova Italy, and }
\centerline{\it LDFC, Institut de Physique,}
\centerline{\it Universit\'e Louis Pasteur, 67084 Strasbourg France}
\centerline{e-mail:  tomaso@ldfc.u-strasbg.fr}
\centerline{\today}
\vskip.7cm
\centerline{\bf Abstract}
\noindent
{\it 
Topologically stable cellular partitions of $D$ dimensional spaces  are studied.
A complete statistical description of the average structural  properties 
of such partition is given in term of a sequence of ${D \over 2}-1$ 
(or ${D-1 \over 2}$) variables for $D$ even (or odd). 
These variables are the average coordination
numbers of the $2k$-dimensional polytopes ($2k < D$) which make the cellular structure.
A procedure to built $D$ dimensional space partitions trough cell-division and cell-coalescence transformations is presented.
Classes of structures which are invariant under these transformations are found and
the average properties of such structures are illustrated.
Homogeneous partitions are constructed and compared with the known 
structures obtained by Vorono\"{\i} partitions and sphere packings in high dimensions.
}

\section{\label{Intr} Introduction }

We study topologically-stable division of any dimensional space by cells.
Such systems have minimal incidence numbers. 
Configurations with higher incidence numbers are topologically unstable because they can be splited  into configurations with the minimal incidence numbers by infinitesimal   local transformations.
In the  literature these cellular partitions are known as ``froths'' since in two and three dimensions ($2D$ and $3D$) the soap froth is the archetype of such structures.
A $2D$ froth is a space-filling cellular partition made of irregular polygons 
where on each vertex  are incident three polygons.
A $3D$ froth is a polyhedral partition of space where on each vertex are incident 4 polyhedra.
In general, a $D$-dimensional froth is a partition of space in irregular polytopes, where on each
vertex $D+1$ polytopes are incident.
Cellular structures with minimal incidence numbers always appear  when the space is filled by 
cells without following any special symmetry.
Therefore, froths are the typical structures of any disordered partition of the space in cells.

A broad class of disordered natural and artificial cellular systems have the topological structure of froths \cite{dAT,WR84,Stavans}.
Examples in two dimensions are magnetic domains in garnets films, B\'ernard-Marangoni 
cells in thermal convections, biological tissues, cuts of polycrystalline 
metals and ceramics, emulsions, the subdivision of territory in administrative regions 
or in national states, geological structures and $2D$ soap froth (which is obtained by squeezing the soap foam between two plates) \cite{Atk88,Smith,Stav93}.
In three dimensions, examples are biological cells, polycrystalline metals and ceramics, foams \cite{Atk88,Momba90,Dub97}. 
Moreover, the structure of any packing (of hard spheres or atoms, for example) is the dual of a cellular system (which can be generated, for instance, by using the  Vorono\"{\i} construction \cite{Voronoi} around the centers of the packed elements). 
In general, cellular systems generated by packings elements without the use of any 
specific symmetry have structures which are topologically froths.
It follows that, among the examples of $3D$ froths one can include amorphous metals, 
glasses  and some crystalline structures such as the tetrahedrally closed-packed 
phases (t.c.p.) \cite{SadocBook,OKeeffeBook}.

Froths in spaces with dimensionality higher than 3 are relevant in information theory and signal processing  \cite{WdW60,Sloane}.
Indeed, an information can be associated with a point in an $N$-dimensional space. 
To transmit and recover the information in presence of noise one must put  the points in the $N$-dimensional space, separated by a certain distance which must be larger than the additional noise.
Therefore to each point (information) is associated a finite volume and the entire space is subdivided in cells each one containing one encoded information \cite{WdW60}.
The energy necessary to transmit an information is proportional to the distance of the representing point respect to the origin.
An efficient coding, which minimizes the energy, organizes the volumes associated with the different information in the closest possible packing of similar cells around the origin \cite{Sloane}.

Dense packings of equal cells in high dimensions have also applications in the study of 
analogue-digital converters.
In this case, the space of the continuous  analogical variables is quantized in a system 
of cells and the volume inside each cell is associated with one digital information. 
The quantization error is associated with  the extension of the interface between the cells and with the distance between the center of a cell and its vertices \cite{Sloane}.

High dimensional partition of space has also application in neural networks and 
complex system dynamics \cite{Gard88,GardDerr88,Bru-Toul92,AmitBook,Kurc95}. 
Some relevant properties (such as the storage capacity in neural network and the slow aging dynamics in glasses) are associated with the subdivision of the phase-space in basin of attraction around the stored information or the minima of the energy. 

Two and four dimensional froths (and their duals: triangulations and simplicial decompositions) have relevance in quantum gravity \cite{Christ82,David94,Ambj94,Ambj95,Ambj96,Wheat94}.
Here the continuous space is divided into cells  and the functional integration over all equivalence classes of metrics is replaced with a summation over all the triangulations of the given manifold.

Despite the  broad  variety of system which are topologically froths and the large amount of studies devoted to them 
in the literature, very little is known about the structure of froths in dimensions larger than $D=3$.
Froths are disordered cellular structures where the cells are highly correlated. 
These correlations essentially come from the space-filling condition which locally constraints
the cells to pack without leaving any empty space, and globally constraints the froth to tile a manifold with a given curvature.
In this  paper we study how these local and global conditions determinate the average topological properties 
of the froth structure and we construct froths in any dimension by cell-division and coalescence transformations.
The aim of the present paper is to investigate  the average structural properties of classes of homogeneous partitions 
of high dimensional Euclidean spaces and to give  analytical instruments and methodologys   for the the investigation
of the topological structure of froths in spaces of arbitrary dimensions and curvature.

The plan of the paper is the following:
In Section \ref{S.1}, the hierarchical organization of topologically stable divisions of space in cells is studied; 
In Section \ref{S.2}, we discuss a way to generate or modify $D$-dimensional froths by 
cell-division and cell-coalescence transformations;
In Section \ref{S.3}, the fixed points of such transformations are studied and
the properties of the associated structures are illustrated;
The construction of homogeneous partitions and the comparison of their properties with known structures, is done in 
Section \ref{S.EH}.

\section{\label{S.1} Hierarchy in the cellular structure}

A froth in an arbitrary dimension $D$ is a cellular structure where the incidence numbers 
(i.e.  the average number of elements which are incident on a given lower-dimensional element) 
are fixed by the stability condition at the minimal value. 
The cells of a froth in dimension $D$ are $D$-dimensional irregular polytopes packed together to fill space. 
The boundaries of these cells are made with $(D-1)$-dimensional polytopes
which are bounded by $(D-2)$-dimensional polytopes and so on up to the zero-dimensional elements which
are the vertices.  
For example, a three dimensional froths is made with $3D$ polyhedra (the cells)
which are bounded by $2D$ polygons (the faces) which are bounded by $1D$ elements (the edges) finally
bounded by $0D$ elements (the vertices).  
A characterization of this $D$-dimensional structure can be
given in term of the numbers of $D$-dimensional polytopes which are making the froth, in
term of the average numbers of $(D-1)$-dimensional polytopes making the boundary of a given cell and so on counting the number of polytopes making the boundary of the boundary, etc. 

The boundary of any $k$-dimensional polytope of the froth is also a froth in a $(k-1)$-dimensional elliptic space.  
The $D$--dimensional froth is therefore a graded topological set:  it contains $D$--polytopes, the cells, which are tiling a space which can be Euclidean, elliptic or hyperbolic.  
The boundary of each cell is an elliptic $D-1$ surface which is tiled by a
$(D-1)$-dimensional froth, whose cells are $(D-1)$--polytopes which are the interfaces bounding the original cell and separating it from its topological neighbours.  Each  interface of the
$(D-1)$--froth is an elliptic $(D-2)$--froth of $(D-2)$--polytopes, which are  separating the cells from their neighbours.  
The graded topological set terminates with edges (segments or convex 1--polytopes),  bounded by 2 vertices (or 0--polytopes).

Let denote with $C_k$ the number of $k$-dimensional cells in the froth ( $C_0$ number of vertices, $C_1$ number of edges, $C_2$ number of faces, $C_3$ number of polyhedra ... $C_D$ number of $D$-dimensional cells.).  
Let denote with $\langle n_{i,j} \rangle$ (for $i \le j$) the average number of $i$-dimensional 
cells which are surrounding and making the boundary of a $j$ dimensional cell 
($\langle n_{0,1} \rangle =2$ number of vertices surrounding an edge, $\langle n_{1,2} \rangle$ 
number of edges per faces, etc.).  

The average  froth structure is characterized by the numbers of polytopes $C_i$ and by  the valences $\langle n_{i,j}\rangle$ (with $0 \le i < j \le D$), which are therefore the variables of the problem.
The total number of these variables is $ {1 \over 2 } (D+1)(D+2)$, but they
are related by the Euler equations and constrained by the stability condition.  
In particular, the numbers of elements in the froth $C_i$ are related through the Euler relation: 
\begin{equation}
\sum_{i=0}^D(-1)^iC_i=\chi_D \label{1} \;\;\;,
\end{equation} 
where $\chi_D$ is the Euler--Poincar\'e characteristic which is associated with the space curvature.  
For $D$ even, opposite signs of $\chi_D$ correspond  to spaces with opposite curvature:  
$\chi_{D}>0$ corresponds to an elliptic $D$--dimensional space and $\chi_{D}<0$ corresponds  to an
hyperbolic space.  
On the other hand, for $D$ odd, this relation between the sign of $\chi_D$ and the space curvature does not holds anymore.

An Euler relation is also satisfied for each froth of the graded topological set. 
That gives a set of relations for the quantities $\langle n_{i,j} \rangle$
\begin{equation} 
\sum_{i=0}^{J-1} (-1)^i \langle n_{i,J} \rangle = 1-(-1)^{J} \;\;\;\; \mbox{(with $J=1,2,...,D$). }
\label{3}
\end{equation} 
Here the factor $1-(-1)^{J} = \chi_{J-1}^{(elliptic)}$ is the Euler--Poincar\'e characteristic for the surface of a $J$-dimensional sphere (which is a $(J-1)$-dimensional elliptic space).

The numbers $C_i$ and the averages $\langle n_{i,j} \rangle$ are related by the stability condition
\begin{equation} 
\left(\begin{array}{cc}{D+1-i}\\{j-i}\end{array}\right) C_i = \langle n_{i,j} \rangle C_j \;\;\;\; \mbox{(with $i \le j \le D$),}
\label{2}
\end{equation}  
where $n_{i,i}=1$.
The binomial coefficient in the left-hand side of Eq.(\ref{2}) is an incidence number 
(number of $j$-dimensional polytopes incident on an $i$-dimensional polytope) which is fixed 
at the minimum value by the stability condition 
(there are $(D+1)$ edges and $\big(^{D+1}_{\;\;2} \big)$ faces incident on each vertex, 
$D$ faces incident on each edge, etc.).

By contrast, the coordination numbers $\langle n_{i,j} \rangle $ with $i<j$ are  variables (except for $n_{0,1}$: every edge is 
bounded by two vertices and therefore $n_{0,1}=2$). 
These variables are not all independent and their range of variability is severely 
restricted by relations (\ref{1}), (\ref{3}) and (\ref{2}) 
(for example, in a $2D$ infinite Euclidean froth we have $\langle n_{1,2}\rangle = \langle n_{0,2}\rangle =6$ as consequence of 
the Euler relation).

One can show that \cite{AsRi} a complete topological characterization of the average structure of a $D$--froth is given by a set of ${D \over 2}-1$ (or ${D-1 \over 2}$) for $D$ even (or odd), independent variables: the even ``valences'' $X_{2l}=\langle n_{2l-1,2l} \rangle$ with 
$2l < D$. 
These valences are the {\it average coordination numbers} (average number of neighbours) 
of the $2l$--dimensional polytopes in the froth.
These are {\it   free variables}. 
On the contrary, the coordination numbers for the odd dimensional polytopes (the odd valences) 
are given in terms of the even valences by the relations
\begin{equation}
\frac{X_1X_2 \cdots X_{J}}{(J)!} - \frac{X_2 \cdots X_{J}}{(J-1)!} + \cdots 
-(-1)^J\frac{X_{J-1}X_{J}}{2} + (-1)^J X_{J} = 1-(-1)^J   \;\;\; \mbox{(for $J =1,2,....D$)}  .
\label{6} 
\end{equation}  
(Which is obtained from Eq.(\ref{3}) associated with Eq.(\ref{2}) and by using the definition $X_k = \langle n_{k-1,k}\rangle$.)
For $J=2l+1$ odd, the left hand term in Eq.(\ref{6}) is equal to 2 and Eq.(\ref{6}) fixes the value of the odd valence $X_{2l+1}$ in terms of the even ones. 
When $J=2l$ even, the left hand term is zero and therefore the even valences $X_{2l}$ are 
free variables.

In $k$ dimensions the polytope with minimal coordination number is a simplex with $k+1$
neighbours. 
Therefore, the valences must stay into the range $k+1 \le X_{k} < \infty$.
The average structure of a $D$-dimensional froth is characterized by a sequence of 
even free valences $\{X_2, X_4, X_6 .....\}$.
For any given sequence of even valences $\{ X_{2l} \}$ the odd ones can be calculated by 
using equations (\ref{6}). 
But only sequences which generate odd valences with $X_{2l+1} \ge 2l+2$ are admissible.
This condition strongly constraint the accessible values of the even valences.

The Euler's formula (\ref{1}), associated with Eq.(\ref{6}) and Eq.(\ref{2}), gives an 
additional  relation between  valences
\begin{equation}
\Bigg( \frac{X_1X_2 \cdots X_{D}}{(D+1)!} - \frac{X_2 \cdots X_{D}}{(D)!} 
+ \cdots -(-1)^D \frac{X_{D}}{2} +(-1)^D \Bigg) C_D = \chi_D \; .
\label{15b} 
\end{equation}

In even dimensional spaces, the sign of the Euler--Poincar\'e characteristic $\chi_D$ is associated 
with the space-curvature. 
The sign of the term inside the brackets in the left-hand side of Eq.(\ref{15b}) is the same 
of $\chi_{D}$ (because $C_{D}>0$).
Therefore, any two regions of the valences' space which have different signs of the bracket term 
(i.e. of the $\chi_{D}$) correspond to two froths on manifolds of opposite Gaussian curvature. 
For example, in the $2D$ case (where the average structure of the froths is described by the 
parameter $X_2$) equation (\ref{15b}) gives 
\begin{equation}
(6-X_2) {C_2 \over 2} = \chi_2 \;\;\;,
\label{2Dcurv} 
\end{equation}
which indicates that $2D$ froths with $X_2<6$ are tiling elliptic surfaces, 
whereas froths with $X_2>6$ are tiling hyperbolic surfaces (see fig.(1a)). 
In four dimensions Eq.(\ref{15b}) gives
\begin{equation}
X_4 = \Big(1-{\displaystyle \chi_4 \over C_4} \Big) 5 {\displaystyle 6- X_2  \over 5-X_2 } \;\;  ,
\label{X4}
\end{equation}
(where we used Eq.(\ref{6}) to express $X_3$ as a function of $X_2$).
Equation (\ref{X4}) indicates that the region in the parameter space $\{ X_2,X_4 \}$ below the 
line $X_4=5{6-X_2 \over 5-X_2}$ is associated with $4D$ froths which are tiling elliptic manifolds ($\chi_4 > 0$), whereas the region above this line correspond to froths tiling hyperbolic manifolds ($\chi_4 < 0$) (see fig.(1b)).

Note that Eq.(\ref{6}) is a constraint on the sequence $\{ X_{2l} \}$ given by local conditions (it
concerns the average properties of a $2l+1$ dimensional polytope in terms of the properties of the lower
dimensional elements that are making it).  
Whereas, Eq.(\ref{15b}) is a constraint on $\{ X_{2l} \}$
given by the global curvature of the manifold that the froth is tiling.

Note also that relation (\ref{X4}) has a singularity in $X_2 =5$.
This is associated with the existence of polytope $\{ 5,3,..\}$ (Schl\"alfly symbols \cite{Cox61}) up to $D=4$ only, as explained in Appendix \ref{A.533}.

\section{\label{S.2} Cell-division and cell-coalescence transformations}

In this section we build $D$-dimensional froths by using the cell-division transformation and its inverse (the cell-coalescence).
This is a local transformation that changes the structure of the froth but leave unchanged the 
global topological properties (the curvature of the manifold tiled by the froth or  -equivalently- 
the parameter $\chi_D$). 
By using cell division and coalescence it is therefore possible to generate different 
froths which are tiling topologically identical manifolds.

In the literature analogous transformations have been studied for the dual problem of 
triangulations and simplicial decomposition. 
In particular, it is known that, given two different partitions of the $D$-dimensional 
space in $D$-simplices (where a 0-simplex is a point, 1-simplex an edge, 2-simplex a triangle, 3-simplex a tetrahedron, etc.), one can be transformed into the other by a finite sequence of two local transformations called ``Alexander moves'' \cite{Alexander}. 
The first move is the addition of a vertex inside a simplex dividing it in $D+1$ simplices 
with the same boundary of the original simplex. 
The second move correspond to add a vertex on an edge of a simplex and connecting it with the 
vertices  of the incident simplexes.
For the $2D$ case the two Alexander move are the insertion of a new vertex inside a triangle 
and the insertion of a new vertex on an existing edge.
They are shown in fig.(2a) and (2b). 
In the dual froth these moves corresponds to a cell-division which inserts a triangle near 
to an existing vertex and to a cell-division which inserts a square near to an existing edge
(fig.(2c) and (2d)).
In $3D$, one can easily see, by following the same procedure illustrated for $2D$, that the 
two Alexander moves can be obtained in the dual space of the froth by 
applying cell-division transformations.
In the general case, one can see that the first Alexander move can be always done in the dual 
froth by dividing a cell in the proximity of a vertex inserting in this way a new polytope 
with $D+1$ neighbours (a simplex). 
The second move can be done dividing a cell in the proximity of an existing
$D-1$ dimensional interface between two cells.
In this case, the kind of polytope inserted depends on the local configuration.

We have therefore shown that, the two Alexander moves are reduced in froths to two special kinds of cell-division 
transformations.
Consequently, {\it  the entire  set of all the possible froths tiling a given manifold can be generated by cell-division and
its inverse (cell-coalescence)  transformations}.

Now we investigate how the average properties of the structure are modified by these transformations.
First consider the cell-division transformation in the two dimensional case.
The cut of a cell corresponds to insert in the system 1 additional face, 3 edges and 2 vertices. 
Therefore, one has the transformations $C_2 \rightarrow C_2+1$, $C_1 \rightarrow C_1+3$ and 
$C_0 \rightarrow C_0+2$. 
One can verify that the Euler-Poincar\'e characteristic rests unchanged (indeed, $\chi_D=C_0-C_1+C_2$). 
On the contrary, the average coordination number (which is given by $X_2={2C_1 \over C_2}$, 
see Eq.(\ref{2})) is modified 
\begin{equation}
X_2'=X_2 \mp {\displaystyle 1 \over C_2 \pm 1} (X_2 - 6) \;\;\;\; ,
\label{one}
\end{equation}
with the upper sign corresponding to a cell-division transformation and the lower 
sign to its inverse (coalescence).

Now consider the $3D$ case. 
A cell-division  corresponds to insert inside a cell  a new face which can have, 
in general, $c_1$ edges and $c_0(=c_1)$ vertices.
This cut corresponds to the transformation $C_3 \rightarrow C_3+1$, $C_2 \rightarrow C_2+1+c_1$, 
$C_1 \rightarrow C_1+c_1+c_0$ and $C_0 \rightarrow C_0+c_0$.
Equation (\ref{2}) gives $X_3={2C_2 \over C_3}$, and therefore we obtain that the average 
coordination number transforms as
\begin{equation}
X_3'=X_3 \mp {\displaystyle 1 \over C_3 \pm 1} \Bigl(X_3-2(c_1+1) \Bigr)\; . 
\label{three}
\end{equation}

In the $D$-dimensional case, the cut of a cell corresponds to introduce a $D-1$ dimensional 
interface which is, in general, made of $c_0$ vertices, $c_1$ edges, $c_2$ faces, $c_3$ 3-dimensional 
cells .... $c_{D-2}$  $(D-2)$-dimensional polytopes. 
Consequently, the division of a $D$-dimensional cell (or the coalescence between two cells) of the 
$D$-dimensional froth correspond to the transformation
\begin{eqnarray}
C_0 &\rightarrow & C_0 \pm c_0 \nonumber \\
C_1 &\rightarrow &  C_1 \pm c_1 \pm c_0 \nonumber \\
C_2 &\rightarrow & C_2 \pm c_2 \pm c_1 \nonumber \\
&\vdots & \nonumber \\
C_k &\rightarrow & C_k \pm c_k \pm c_{k-1} \nonumber \\
&\vdots &\nonumber \\
C_{D-1} &\rightarrow & C_{D-1} \pm 1 \pm c_{D-2} \nonumber \\
C_D & \rightarrow & C_D \pm 1 \;\;\;,
\label{transf}
\end{eqnarray}
where  the upper sign $(+)$ corresponds to a cell-division transformation and the lower 
sign $(-)$ to its inverse (coalescence).
By substituting into Eq.(\ref{1}) one can verify that the global curvature ($\chi_D$) is an 
invariant quantity under the transformation (\ref{transf}).
Note that, expression (\ref{transf}) takes the canonical form $C_k \rightarrow C_k \pm c_k \pm c_{k-1}$ for all $k$ if one impose $c_{-1}=0$, $c_{D-1}=1$ and $c_D=0$.

From Eq.(\ref{2}) one has the identity $X_k C_k = (D+2 -k) C_{k-1}$ (where we used the definition 
$X_k = \langle n_{k-1,k} \rangle$).
By substituting in this expression the transformation (\ref{transf}) we get
\begin{equation}
X'_k = X_k \mp {\displaystyle 1 \over C_k \pm c_k \pm c_{k-1} }
                 \Big( X_k ( c_k +c_{k-1}) - (D+2-k) (c_{k-1} + c_{k-2}) \Big) \;\;,
\label{X'}
\end{equation}
(upper sign, cell-division; lower sign, cell-coalescence).

We recall that through cell-division/coalescence transformations  it is possible to generate the 
full class of froths tiling topologically identical manifolds.
The modification of the average structural properties associated with these geometrical 
transformation are algebraically given by equation (\ref{X'}). 
By using this expression it is therefore possible to {\it  find the average topological  properties of 
all the froths on a given manifold}.

\section{\label{S.3} Fixed points}

When cell-division or coalescence transformations are performed on a 
$2D$ froth with  average coordination $X_2=X^*_2 =  6$,  they leave  the local average structural properties 
unchanged  (i.e. $X'_2=X_2=X^*_2 =  6$,  see Eq.(\ref{one})).
This is a fixed point in the transformation (\ref{transf}) and corresponds to $2D$ Euclidean froths.
Moreover, one can see that the average structural properties of froths which are tiling 
elliptic surfaces (i.e. the one with $X_2<6$) are modified toward the Euclidean structure 
($X_2 < X'_2 <6$) by the application of the cell division transformation. 
Analogously hyperbolic froths ($X_2>6$) are also modified towards the Euclidean structure 
($6<X'_2 < X_2$) (see fig.(1a)). 
(Note that the global curvature remain always unchanged. 
Indeed, $\chi_D$ is an invariant under the transformation (\ref{transf})).

In the general case, one can immediately see that transformation (\ref{X'}) has the fixed point
\begin{equation}
X_k^* =  (D+2-k){\displaystyle c_{k-1} + c_{k-2} \over c_k + c_{k-1} } \;\;\;,
\label{f.p.0}
\end{equation}
which is the structure that is invariant under cell-division/coalescence transformations ($ \{ {X^*_k}' \} = \{ X^*_k\} $). 

A froth is a graded set. 
Therefore the $D-1$ dimensional interface that have been 
introduced into the system to cut a cell is a $(D-2)$-dimensional elliptic froth with $c_0$ vertices, 
$c_1$ edges .... $c_{D-2}$ $(D-2)$-dimensional cells. 
All the relations written above, and in particular Eqs.(\ref{3}) and (\ref{2}), can be applied to 
this $(D-2)$-dimensional elliptic froth. 
One has, $c_k x_k= (D-k) c_{k-1}$ and $c_{k-1} x_{k-1} = (D+1-k) c_{k-2}$, with $x_k$ and $x_{k-1}$ 
the average coordination numbers of the $k$ and $(k-1)$-dimensional polytopes which are making the 
$(D-1)$-dimensional interface.
By substituting into Eq.(\ref{f.p.0}), one gets
\begin{equation}
X_k^* = {\displaystyle (D+2-k)(D+1-k+x_{k-1}) \over (D+1-k)(D-k+x_k) } x_k \;\;\;.
\label{f.p.}
\end{equation}
The fixed point configuration is therefore determined by a set of variables $\{ x_k \}$ with $k < D$
which are the average coordinations of the ($D-1$)-dimensional polytope that is inserted or removed 
during the cell division or coalescence transformation.
For example, in $D=3$, relation (\ref{f.p.}) gives
\begin{equation}
X^*_3 = 2 x_2 +2
\label{f.p.3D}
\end{equation}
with $x_2$ the number of edges of the face that is inserted (or removed) to divide a cell (or make coalescence between two cells).

\begin{figure}
\vspace{-2cm}
\epsfxsize=10.cm
\epsffile{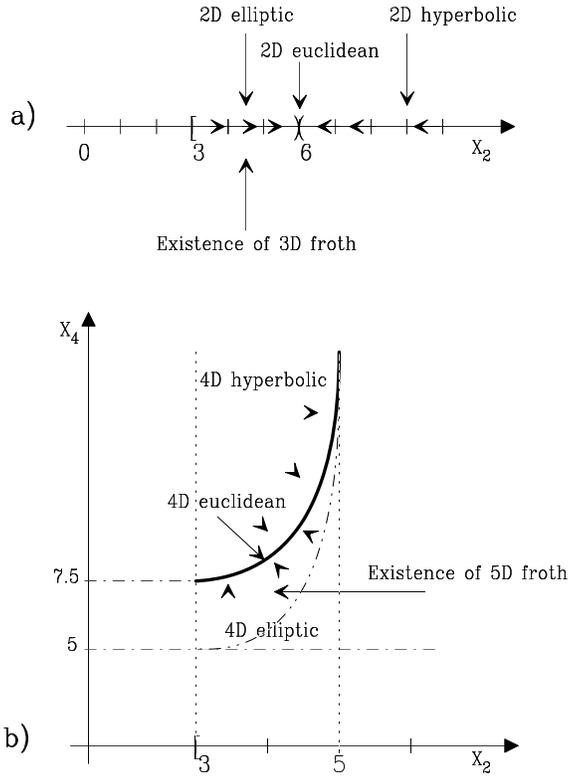}
\begin{centering}
\vspace{0cm}
\caption{
\label{f.1} 
Different regions in the parameter space $\{ X_{2l} \}$ corresponds to froths which are tiling spaces of different curvatures. {\bf (a)} two dimensional froths with $X_2 < 6$ tile elliptic surfaces, froths with $X_2 > 6$ tile hyperbolic surfaces and $X_2=6$ corresponds to froths tiling the Euclidean plane. {\bf (b)} The hyperbolic, Euclidean and elliptic tilings corresponds to three regions of the $\{X_2,X_4\}$ parameter space.
Cell division transformations modify the properties of curved tilings towards the Euclidean ones (arrows).
}
\end{centering}
\end{figure}

The minimum number of edges per cell is 3. 
Therefore from Eq.(\ref{f.p.3D}) follows that fixed point structures are possible only in the region of the parameter space with $X_3 \ge 8$ (see fig.(\ref{f.1} a)).
Any structure with $X_3 <  8$, is transformed towards the fixed point region ($X_3 \ge 8$) by applying cell-division transformations.

In four dimensions Eq.(\ref{f.p.}) gives
\begin{eqnarray}
X^*_2 &=& {\displaystyle 20 x_2 \over 3(2+x_2) } \nonumber \\
X^*_4 &=& 2 (x_3 +1) = 2 \Big( {\displaystyle 12 \over 6- x_2 } + 1 \Big)\;.
\label{f.p.4}
\end{eqnarray}

We can express the parameter $x_2$ in Eq.(\ref{f.p.4}) in terms of $X^*_2$ obtaining 
$X^*_4 = 5 {6- X^*_2  \over 5-X^*_2 }$,  which is the condition  on the even valences that 
identifies the Euclidean region in 4-dimensional froths (see Eq.(\ref{X4})).
{\it  The fixed point structures are Euclidean}.
Since $x_3 \ge 4$, follows $X_4^* \ge 10$, which implies that structures generated by cell division can only access to 
a part of the Euclidean region in the phase-space.

We can in general prove that the fixed point $\{ X^*_k \}$ given by Eq.(\ref{f.p.}), 
is the average structure of a $D$-dimensional froth which is tiling a manifold with $\chi_D=0$. 
Indeed, let substitute into Eq.(\ref{15b}) the fixed point configuration ($\{ X^*_k\}$) and 
apply the cell division transformation. 
By definition the sequence $\{ X^*_k\}$ doesn't change, whereas the total number of cells increases of a
unity ($C_d \rightarrow C_D +1$).
To satisfy Eq.(\ref{15b}) before and after this transformation one must have $\chi_D =0$.
Which prove the theorem.

By re-writing Eq.(\ref{X'}) in the form
\begin{equation}
X'_k = X_k \mp {\displaystyle ( c_k +c_{k-1}) \over C_k \pm c_k \pm c_{k-1} }
                 \Big( X_k  - X^*_k \Big) \;\;,
\label{X'X*}
\end{equation}
it is easy to see that the fixed points are stable under cell division transformations (upper sign in
Eq.(\ref{X'X*})) which inserts identical polytopes as interface.
Indeed, from Eq.(\ref{X'X*}), if $X_k > X^*_k$ then $X_k > X'_k > X_k^*$ and vice versa.

In the space of the configuration \{~$X_2, X_4, ...,X_{2l},...$~\}, when $D$ even, the Euclidean region
is a surface given by Eq.(\ref{15b}) (with $\chi_D =0$).
The fixed point configurations are  a subset of this surface. 
Froths outside the fixed point configuration are always transformed toward this subset by 
applying   cell-division transformations.
When $D$ odd, Eq.(\ref{15b}) is not a constraint and the fixed point are associated with a sub-volume of the whole accessible parameter space.

\section{\label{S.EH} Construction of Euclidean froths}

In this paragraph we study froths generated by cell-division transformations.
We therefore study the class of structures $\{ X^*_k \}$ given by Eqs.(\ref{X'}) and (\ref{f.p.}). 
The full class of these froths  is obtained by varying in Eq.(\ref{f.p.}) the parameters 
$\{x_k \}$ in the allowed range (i.e. $k+1 \le x_k < \infty$, which satisfy the conditions 
(\ref{6}) for $k$ odd and the relation (\ref{15b}) with $\chi_D<0$).
Here, we study only some particular cases.

Let us  first note that, from Eq.(\ref{f.p.}), the average number of neighbours of the fixed point structure is given in term of the coordination of the inserted interface ($x_{D-1}$) by
\begin{equation}
X^*_D = 2 x_{D-1} +2 \;\;\;\;.
\label{coord}
\end{equation}

In two dimensions an edge is inserted or removed from a face.
The ``coordination'' of an edge is its number of vertices:  $x_{D-1}=x_1=2$.
Therefore $X^*_2 = 6$, as should be in Euclidean space.
In three dimensions a face is inserted in, or removed from a cell.
The coordination of this face ($x_2$) is its number of edges and in principle it can be any number 
between 3 and $\infty$.
But,  a face with a large number of edges can be inserted only in a cell with a large number of neighbours and it can be removed  only if it exists in the froth.
Therefore, only some values of $x_2$ are admissible.
One can easily see that a triangle ($x_2=3$) can always be inserted in the proximity of a vertex.
Analogously, a square ($x_2=4$) can also be always inserted in the proximity of an edge. 
From (\ref{coord}) follows therefore that  three dimensional Euclidean froths with $8 \le X^*_3 \le 10$ 
can always be generated. 
But, in general, it should be possible to insert faces with higher values of $x_2$.
To have an estimation for the ``typical'' value for the number of edges of the inserted face let make
a cut of the whole three dimensional froth with a plane.
The result is a two dimensional Euclidean froth where each single face is the result of a cut on a three dimensional cell.
This two dimensional froth is therefore a representative set of faces produced by random cuts of 
three dimensional cells.
The average number of edges for this set of faces is $x_2 = X^*_2 = 6$.
Therefore, from (\ref{coord}),  a ``typical''  froth generated by cell division is expected to have a fixed 
point coordination around $X^*_3 = 14$ \cite{Lews}.
Cells in biological tissues appear in various polyhedral shapes with a number of faces distributed in a narrow
range around 14 \cite{Dorm}.
A widely studied three dimensional froth made with identical cells is the ``Kelvin froth'', 
its cells are space-filling truncated octahedra with $X_3 =14$ \cite{Kelvin,KelvinBook}.
Coordinations between 15.53 and 14  are found in  Vorono\"{\i} partition of space \cite{Oger96}, 
where the higher value corresponds to a  Vorono\"{\i} partition  from random points \cite{Mei53}
whereas the lower value corresponds to more compact and homogeneous packings .
Smaller values in the range $13.333 \le X^*_3 \le 13.5$ characterize an interesting class of natural structures 
(Frank-Kasper phases \cite{FK}) which partitionate the ordinary space with cells with pentagonal and 
hexagonal faces only. 
Soap froth has typically $X_3 \simeq  13.7$ \cite{Matz}.

In a $D$-dimensional froth a ($D-1$)-dimensional interface with coordination $x_{D-1}$ is inserted or removed
by cell division or coalescence transformations.
As pointed out above, simplexes with coordinations $x_{D-1} =D$ can always be inserted in the proximity
of an existing vertex.
This is the minimum possible value for $x_{D-1}$ and substituted into Eq.(\ref{coord}) sets the minimum value of the average number of neighbours 
in a $D$-dimensional Euclidean fixed point structure at the {\it minimal coordination  }
\begin{equation}
 X^*_D = 2D +2 \;\;\;.
\end{equation}

The argument for the ``typical'' cut that we used in three dimensions can be directly extended to any dimension.
Indeed, one of the properties of froths is that a cut with an hyper-plane of a $D$-dimensional froth 
generates a $D-1$-dimensional Euclidean froth. 
For instance, a cut of a four dimensional froth give a three dimensional Euclidean froth.
We can assume, that this froth has the ``typical'' coordination $X^*_3 =14$ found above.
Inserting into Eq.(\ref{coord}) one gets $X^*_4 = 30$.
The same arguments, extended to any dimension, give 
\begin{equation}
 X^*_D = 2^{(D+1)}-2\;\;\;\;\;,
\end{equation} 
for the average number of neighbours per cell in the ``typical'' $D$-dimensional froth.

What makes  Eq.(\ref{f.p.}) powerful is the fact that,  not only the average number of neighbours,  but {\it all the average 
properties of the fixed point structures can be deduced in term of the properties of the inserted interface}.

\subsection{Minimally coordinated froths}

Let first construct the Euclidean froth with minimum coordination numbers.
It is the fixed point structure associated to a cell-division transformation which inserts interfaces with minimal coordinations. 
This interfaces are $D-1$ dimensional simplices inserted in the proximity of vertex.
They have $x_k = k+1$ (for $k < D$).
By substituting into Eq.(\ref{f.p.}) one obtains
\begin{equation}
X^*_k={\displaystyle D+2-k \over D+1-k}(k+1)\;\;\;.
\label{k+1}
\end{equation}
(Note that $X^*_D = 2D+2$, as discussed above.)
This are the average structural properties of a froth which is tiling a manifold with $\chi_D = 0$ which is 
homologue to the Euclidean space. 
It is the known Euclidean forth with minimal coordination numbers.
Starting from any given $D$-dimensional froth one can always transform it into this minimally
coordinated one by applying an infinite number of cell-divisions near existing vertices.
The resulting structure is expected to have cells with very different topological properties. 
Indeed, for each cell-division transformation a new cell with $D+1$ neighbours is inserted and 1 
neighbour is added to the $D+1$ cells around the inserted simplex,
distributing therefore the coordinations inhomogeneously between cells.

\subsection{Homogeneous partitions}

A $D$-dimensional froth has $D+1$ edges incident on each vertex.
In an ideal homogeneous partition of space these edges are equally separated in angle.
That corresponds to an angle  $\theta^{ideal} = \cos^{-1}(-1/D)$ between each 
couple of edges \cite{AsRi}.
In a froth, edges must close in rings which are bounding two dimensional faces.
It is easy to see that with the angle $\theta^{ideal}$, flat rings close with an average number of edges equal to:
\begin{equation}
X_2^{ideal} = { 2 \pi \over \pi - \cos^{-1}\big(-{ 1 \over D}\big)}\;\;\;.
\label{X2id}
\end{equation}
This number is 6 in two dimension, 5.104 in three dimensions, 4.767 for $D=4$, 4.588 for $D=5$
and tends to 4 when $D \rightarrow \infty$.
Note that $X^{ideal}_2$ is irrational for $D>2$. 
In the Euclidean space, the ``ideal'' structure cannot be obtained by any ordered lattice structure. 
This is an example of geometrical frustration.
But disordered or non-periodic structures can approximate $X_2^{ideal}$ with arbitrary precision avoiding in this way the frustration. 

The average number of edges per cell $X_2^{ideal}$ (Eq.(\ref{X2id})) is the only quantity, of the ideal structure, that can be calculated by using these geometrical arguments.
All the other coordinations are unknown, but we can construct fixed point structures that
approximate this ideal froth in the ring coordination $X_2$.
For these   structures we can calculate the whole set of coordinations, and 
therefore we can infer information about the coordinations of the ideal one.

We expect that structures that partitionate uniformly space must have $X_2 \simeq X^{ideal}_2$.
By cell division transformation it is possible to generate froths that approximate the ideal structures by inserting interfaces with $x_2$ close to the ideal value $X^{ideal}_2$.

For $D=3$,  $X^{ideal}_2 = 5.104$ which corresponds to $X_3^{ideal} = 13.392$. 
In $3D$ we can generate homogeneous partitions by inserting pentagons $x_2=5$ or hexagons  $x_2=6$ obtaining (from Eq.(\ref{coord})) $12 \le X^*_3 \le 14$, which is in the right range. 

In general, since we are looking for homogeneity, it is logical to insert as interfaces 
regular polytopes with $x_2$ close to the ideal value $X^{ideal}_2 $.
These polytopes can only be hyper-cubes \{4,3...\}  (which have $x_2 =4$) 
and the polytope \{5,3...\} (with  $x_2 =5$), but this second polytope exists only
up to $D=4$ (\cite{Cox61} and  App. \ref{A.533}).

\bigskip 
\noindent
{\bf (a)}
Cell-division operations which inserts the polytope \{5,3...\} can  therefore generate Euclidean fixed point structures up to $D=5$.
In $D=3$ this corresponds to  cell-divisions  which   insert   pentagonal faces,
that generates a fixed point structure with $X^*_2 = 5$ and  $X_3^* = 12$. 
In four dimensions, the fixed point structure obtained by inserting   dodecahedra 
(\{5,3\}, $x_2 = 5$, $x_3 = 12$) has $X^*_2 = 100/21 = 4.7619...$ value that is very close to the ideal one.
The four dimensional cells of this froth have $X^*_4 =26$ neighbours in average.
For $D=5$, by dividing cells with  the polytope \{5,3,3 \} ($x_2 = 5$, $x_3 = 12$, $x_4 = 120$) 
we obtain a fixed point structure with  $X^*_2 = 75/16 = 4.68...$  (see Eq.(\ref{f.p.})) which is
larger than the ideal value.
The average number of neighbours is in this case $X^*_5 = 242$.

\bigskip
\noindent
{\bf (b)} 
The average coordinations of the fixed point froths  generated by inserting hyper-cubes \{4,3,...\}  are given by imposing 
$x_k = 2k$ into Eq.(\ref{s.a}))

\begin{equation}
\mbox{\large \bf (b) }  \;\;\;\;\;\;
X^*_k={\displaystyle (D+2-k)(D-1+k) \over (D+1-k)(D+k)}2k\;\;\;.
\label{s.a}
\end{equation}
In this structure the $D$-dimensional cells have $X_D^* = 4D-2$ neighbours in average.
The average ring-coordination is $X^*_2 = 4 D(D+1)[(D-1)(D+2)]^{-1}$ which correctly tends to 4 when $D \rightarrow \infty$, but it is systematically lower than $X^{ideal}_2$ for $D>2$.
This is presumably a rather inhomogeneous structure

\bigskip
\noindent
{\bf (c)} 
To maximize homogeneity one can construct a structure by inserting new interfaces with the same 
topological properties of the existing structure. 
We expect a resulting structure that evolves toward a self-uniform homogeneous partition. 
Let therefore perform cell-division transformations by inserting interfaces with 
$x_k = X^*_k$ (with $k=2, ... ,D-1$). 
By substituting into Eq.(\ref{f.p.}) one obtains a recursive equation with the 
following solution
\begin{equation}
\mbox{\large \bf (c) } \;\;\;\;\;\;
X^*_k=\Big( 1+ {\displaystyle D \over D+1-k} \Big) k  \;\;\;.
\label{s.b}
\end{equation}
Here $X_d = D(D+1)$ and $X_2 = 2 {2D -1 \over D-1}$, which asymptotically tends to 4 and is much closer to the ideal value than the one of structure  (b).

\bigskip
\noindent
{\bf (d)} 
Partitions can be generated by inserting ``typical'' interfaces as described before.
Here the ``typical'' ($D-1$)-dimensional interface has coordinations $x_k$ which are equal to average coordinations 
of the fixed point Euclidean structure obtained with this procedure in $D-1$ dimensions.
The values $X^*_k$ are then given in term of a recursive equation (with initial condition $X^*_1 = 2$).
Here are the solutions  for $k=2$ and $k=D$, which have a simple compact form
\begin{eqnarray}
\mbox{\large \bf (d) }  \;\;\;\;\;\;
X^*_2 & = & 12 { D \over 3D -2} \nonumber \\
X^*_D & = & 2^{(D+1)}-2 \;\;\;\;.
\label{s.c}
\end{eqnarray}
Surprisingly the product of these valences from $k=1$ to $D$ has also a very simple form:
$X^*_1 X^*_2 X^*_3 \cdots X^*_D = D!(D+1)!$. 
Here, the value of $X_2$ is systematically bigger than the one of the ideal structure but it is extremely close to it.

In table 1 the values of $X_2$ and $X_D$ are reported, up to $D=10$, for the whole set of fixed point structures which have been studied in this paragraph.
In Fig.(\ref{f.X2}) the value of $X_2$ for the ideal partition and for the fixed point structures (a), (b), (c), (d) are plotted up to $D=24$.

\subsection{Kissing Numbers}

The average number of neighbours $X_D$  of a $D$-dimensional cell in an  Euclidean froth is an interesting quantity.
In sphere packings a corresponding quantity  is called ``kissing number'',  it is the number of identical spheres that
can be placed around a given sphere being in contact (kissing) with it  \cite{Sloane}.
To find sphere packings configurations with high kissing numbers has relevance in the design of efficient codes.
It is known that, for packings of identical  spheres, the kissing number (KN) is 6 in $D=2$, 12 in $D=3$, 
but exact answers are unknown for dimensions above 3 except for $D=8$ (KN=240) and $D=24$ (KN=196560) 
where two specially dense lattices $E_8$ and $A_{24}$ achieve the maximal possible values of  KN. 
In fig.(\ref{f.kiss}) are reported the values of the highest known kissing numbers for lattice and non-lattice sphere packings.
Two known bounds for KN when $D \gg 1$ are also reported.
The kissing number question concerns to find the best local arrangements of spheres.
In high dimensional spaces, this configuration does not necessarily corresponds to any lattice packing.
Disordered or quasi-ordered packings are often more suitable to attain high kissing numbers.  
Dimension $D=9$ is the first where non-lattice packings are known to be superior.
Here the Leech lattice $\Lambda_9$ has KN=272 whereas the best bound known is 380 \cite{Sloane}.

To any sphere packing one can associate a cellular structure constructed by partitioning the space in convex polytopes 
each one containing inside a sphere.
Kissing spheres are neighbours. 
In a dense sphere packing the  enveloping polytopes make a space-filling partition of space.
The number of neighbours of this system of  polytopes is related with the kissing number and
it is expected to be bigger than KN, because some non-kissing spheres can be first neighbours in the associated froth.
This is for instance the case in $D=3$ where the configuration with KN=12 corresponds to a close packing of 
spheres with  an associated Wigner Seitz cell that do not pack  in a froth:  
the incidence numbers  are not minimal.
This  is a {\it topologically unstable} configuration.
Infinitesimal random displacements change the number of  topological neighbours from 12 to an average value of 14, 
but in this case neighbouring spheres will be not all in contact.
In general,  in close packings, we expect the number of neighbours $X_D$ of the enveloping polytopes to be bigger, 
but of the same order of magnitude, of  the kissing numbers of the enveloped spheres. 
 
In fig.(\ref{f.kiss}) the kissing numbers for some known sphere packings are compared with the coordination numbers obtained from our homogeneous partitions (a), (b), (c) and (d) up to $D=24$.

\subsection{Vorono\"{\i} partitions}

The average number of vertices on the boundary of a $D$-dimensional cell ($\langle n_{0,D} \rangle$) can be exactly 
calculated for Vorono\"{\i} partitions generated from random points \cite{Mei53,ItzDru88}:
\begin{equation}
\langle n_{0,D} \rangle = {2 \over D}{\Gamma(D) \over \Gamma\left({1 \over 2}(D+1)\right)^2}
\left[{\Gamma\left({1 \over 2}\right)\Gamma\left({1 \over 2}D+1\right)    \over \Gamma\left({1 \over 2}(D+1) \right)} \right]^{D-1}
{\Gamma\left({1 \over 2}\right)\Gamma\left({1 \over 2}(D^2+1)\right)    \over \Gamma\left({1 \over 2}D^2 \right)}   \;\;.    
\label{Itz}
\end{equation}
Asymptotically this quantity scales as $\langle n_{0,D} \rangle  \propto D^{D/2-1}$.

The average number of vertices per cell can be expressed in term of the coordinations by
\begin{equation}
\langle n_{0,D} \rangle = {1 \over D!} {X_1 X_2 \cdots X_{D-1}X_D }\;\;\;.
\label{nd}
\end{equation}
(Note that for $D=3$ Eq.(\ref{Itz}) give $\langle n_{0,D} \rangle = 96\pi^2/35$  that substituted into Eq.( \ref{nd}) 
leads to  $X_3 = 2+48\pi^2/35=15.53...$).
By substituting the fixed point configuration into (\ref{nd}) we find that the structure  (b) has 
$\langle n_{0,D} \rangle = (D+1) 2^{D-1}$, the structure  (c) gives $\langle n_{0,D} \rangle = (2D)!(D!)^{-2}$, whereas (d) has $\langle n_{0,D} \rangle = (D+1)!$.
In figure (\ref{f.voronoi}) the behaviors of the average number of vertices per cell in the Vorono\"{\i} froth and in the three structures (b), (c) and (d) is shown for $3 \le D \le 50$.

\section{\label{Con} Conclusions}

The topological structure of a $D$-dimensional cellular system can be characterized, in average, in terms of the coordinations ($X_k$) of the irregular polytopes which are making the structure (Section \ref{S.1}).
Only the coordinations of the even dimensional polytopes ($X_{2l}$) are necessary for this characterization, the odd ones are expressed in terms of the even ones by the relation (\ref{6}).
Therefore, the average structure of a $D$-dimensional froth is characterized by a sequence $\{X_2, X_4, X_6 .....\}$ of ${D \over 2}-1$ (or ${D-1 \over 2}$) variables for $D$ even (or odd).   
These variables are related with the space-curvature through Eq.(\ref{15b}). 
Regions in the parameter space $\{ X_{2l} \}$ corresponding to $D$-dimensional froths tiling spaces of different curvature are   discussed for $D \le 4$ (fig(\ref{f.1}) and Eqs.(\ref{2Dcurv}) and (\ref{X4})). 

\bigskip
We use cell division and coalescence transformations to build $D$-dimensional froths (Section \ref{S.2}).
We show that through cell-division/coalescence transformations  it is possible to generate the 
entire class of froths tiling topologically identical manifolds.
The dynamical renormalization of the variables $X_k$ under such transformations is found (Eq.(\ref{X'})). 
The existence of classes of structures which are invariant (fixed points) under cell division/coalescence is pointed out (Eqs.(\ref{f.p.0}) and (\ref{f.p.}) ).
We show that these structures are tiling Euclidean spaces.

\bigskip
Several fixed point Euclidean structures are constructed in Section \ref{S.EH}.
We discuss the average statistical properties for the  of minimally coordinated Euclidean froths, 
and for  several topologically homogeneous space partitions  (Eqs(\ref{k+1}-\ref{s.c}) and tab.(\ref{1})).
The topological properties of the most homogeneous cellular partition are searched   
and compared  with known geometrical results  (fig.(\ref{f.X2})). 

Finally, the fixed point Euclidean structures are compared with known high dimensional structures generated by sphere packings and Vorono\"{\i} constructions (figs.(\ref{f.kiss}) and (\ref{f.voronoi})).

\vskip 20pt
\noindent{\bf Acknowledgements}

\noindent
A special thank to Nicolas Rivier for many discussions and comments which have highly contributed to the research and have improved the presentation of the work.
I acknowledge discussions with J. F. Wheather and D. Sherrington.
This work was partially supported from the European Union Marie Curie fellowship 
(TMR  ERBFMBICT950380).

\appendix
\section{\label{A.533} The existence of polytopes $\{5,3,...\}$ up to $D=4$}

There is a class of regular polytopes with $X_2=5$ and minimally connected vertices ($\{5,3,...\}$) which exists up to dimension $D=4$ \cite{Cox61}.

They are pentagons in $D=2$, dodecahedra ($\{5,3\}$) in $D=3$ and polytopes $\{5,3,3\}$ in $D=4$.
A tessellation of pentagons makes a $2D$ elliptic froth with $X_2=5$ and $C_2=12$, it is a dodecahedron $\{5,3\}$.
A tessellation of dodecahedra make a $3D$ elliptic froths with $X_2= 5$, $X_3=12$, $C_3= 120$, which is the $\{5,3,3\}$ structure.
It turns out that a tessellation with $\{5,3,3\}$ polytopes does not makes any 5-dimensional polytope \cite{Cox61}. 
If existing, such a structure would be a 4-dimensional polytope with  $X_2= 5$, $X_3=12$, $X_4= 120$. 
By substituting these values into Eq.(\ref{X4}), we get $C_4=\chi_4$, which implies $\chi_4 >0$. This hypothetical structure would therefore be an elliptic froth, homotrope to a sphere which implies $\chi_4 = 2$.
Then, from the previous identity, $C_4 =2$.
The hypothetical $\{5,3,3,3\}$ structure would be an elliptic froths which closes onto itself with two cells only.
But two cells are insufficient to make a 5-dimensional polytope (the minimum number is 6).
It follows therefore that the $\{5,3,3,3\}$ structure does not exists and neither exist the others  higher dimensional tessellations  $\{5,3,3,3 ...\}$.

\bibliographystyle{unsrt}

\newpage
\noindent
\begin{centering}
\begin{tabular}{|c | c | c c | c c | c c | c c |}
\hline
  & Ideal &\multicolumn{2}{c|}{FP \{5,3,...\} (a)}&\multicolumn{2}{c|}{FP \{4,3,...\} (b) }&\multicolumn{2}{c|}{FP  (c) } &\multicolumn{2}{c|}{FP  (d) }  \\ 
\hline
   & $X_2^{ideal}$ &  $X_2^*$ & $X_D^* $  &  $X_2^*$ & $X_D^*$ &  $X_2^*$ & $X_D^*$ &  $X_2^*$ & $X_D^*$  \\   
$D$& &  $5{D(D+1) \over (D-1)(D+3)}$ &  $4 {D(D+1) \over (D-1)(D+2)}$ & $\scriptstyle 4D-2$  &
   &  $ 2 {2D -1 \over D-1}$ & $\scriptstyle D(D+1)$ &  $12 { D \over 3D -2}$ & $\scriptstyle 2^{D+1}-2$  \\   
\hline
3  &  {\bf 5.1043}  &{\bf 5  }    & {\bf 12}     & 4.8       & 10             & {\bf 5}   & {\bf 12}  & {\bf 5.1458} & {\bf 14} \\
4  &  {\bf 4.7668}  &{\bf 4.7619} & {\bf 26}     & 4.4444    & 14             & {\bf 4.6667}   & {\bf 20}  &       4.8    &   30     \\
5  &  {\bf 4.5881}  & 4.6875      &  242         & 4.2857  & 18   & {\bf 4.5 }     & {\bf 30}  & {\bf 4.6154 }& {\bf 62} \\
6  &  {\bf 4.7728}  && & 4.2     & 22            & {\bf 4.4 }     & {\bf 42 } & {\bf  4.5 }   & {\bf 126}\\
7  &  {\bf 4.4017}  && & 4.1481  & 26            & {\bf 4.3333 }  & {\bf 56 } & {\bf  4.4211}& {\bf 254}\\
8  &  {\bf 4.3468}  && & 4.1143  & 30            & {\bf 4.2857 }  & {\bf 72 } & {\bf  4.3636}& {\bf 510}\\
9  &  {\bf 4.3052}  && & 4.0909  & 34            & {\bf 4.25   }  & {\bf 90 } & {\bf  4.32  }& {\bf 1022}\\
10 &  {\bf 4.2724}  && & 4.0747  & 38            & {\bf 4.2222 }  & {\bf 110} & {\bf  4.2857}& {\bf 2046}\\
 \hline
\end{tabular}
\end{centering}

\bigskip 
\noindent
Table 1: Average ring coordination ($X_2$) and cell coordination ($X_D$) for some Euclidean fixed point structures (FP) 
generated by cell division (see text).

\newpage

\newpage
\begin{figure}
\vspace{-1cm}
\epsfxsize=14.cm
\epsffile{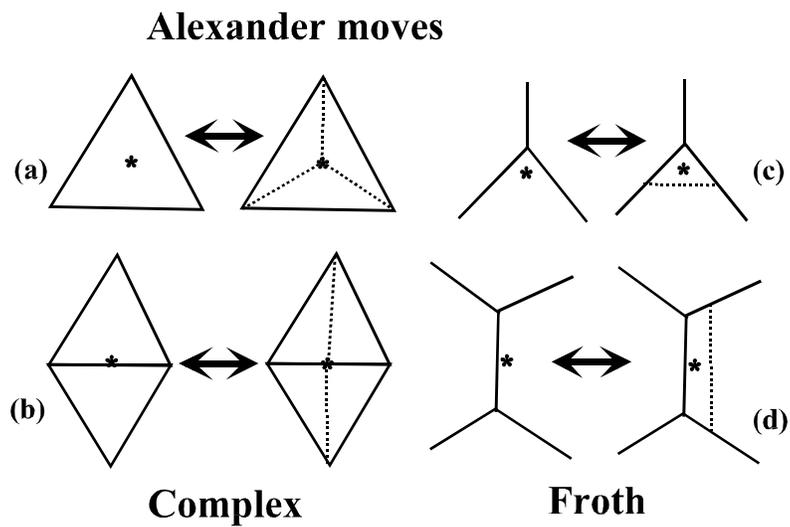}
\begin{centering}
\vspace{-5cm}
\caption{\label{f.2} Given two different partitions of the $D$-dimensional 
space in $D$-simplices one can be transformed into the other by a finite sequence of two local transformations called ``Alexander moves'' ({\bf (a)} and {\bf (b)} for the two dimensional case).
In the dual froths these transformations corresponds to two special cell division transformations ({\bf (c)} and {\bf (d)} for the two dimensional case)}
\end{centering}
\end{figure}

\begin{figure}
\vspace{0cm}
\epsfxsize=14.cm
\epsffile{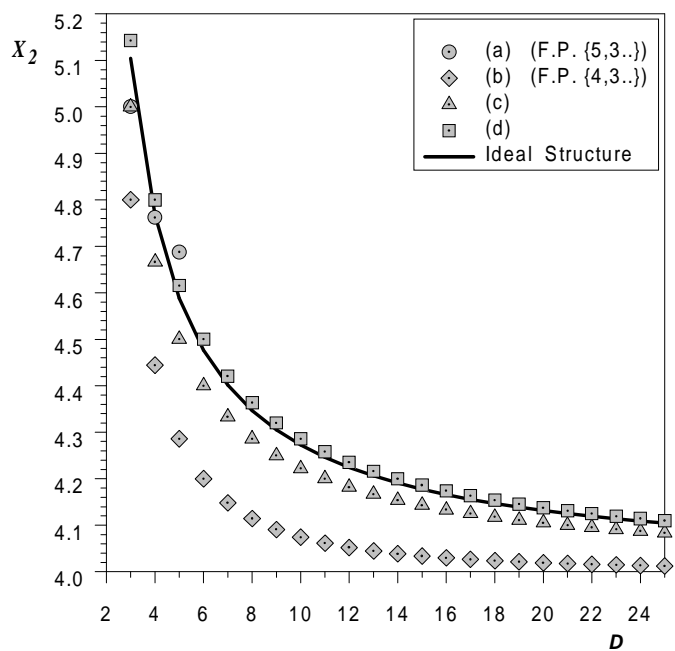}
\begin{centering}
\vspace{-3cm}
\caption{\label{f.X2} The average ring connectivity $X_2$, for homogeneous partitions of spaces of dimension $D$, obtained from a geometrical approach [25] (full line) is compared with the one of the fixed point structures (symbols). }
\end{centering}
\end{figure}

\begin{figure}
\vspace{0cm}
\epsfxsize=14.cm
\epsffile{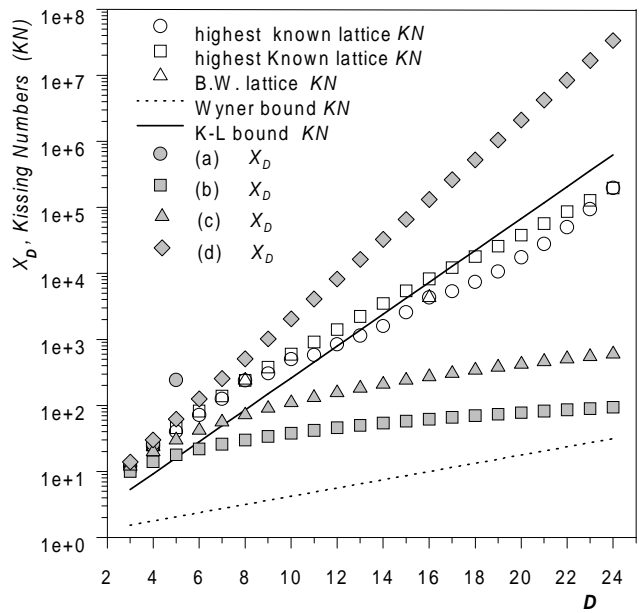}
\begin{centering}
\vspace{-3cm}
\caption{\label{f.kiss} Kissing Numbers KN (maximum number of identical  hard spheres that can stay around and touch a given sphere in $D$ dimensions) and coordination numbers $X_D$ (average number of cells around a given cell in a $D$ dimensional froth) are compared for some known sphere packings (open symbols) and for the fixed point structures (full symbols).  Two upper and lower bounds for the KN in high dimensions are also plotted (full and dotted lines).}
\end{centering}
\end{figure}

\begin{figure}
\vspace{0cm}
\epsfxsize=14.cm
\epsffile{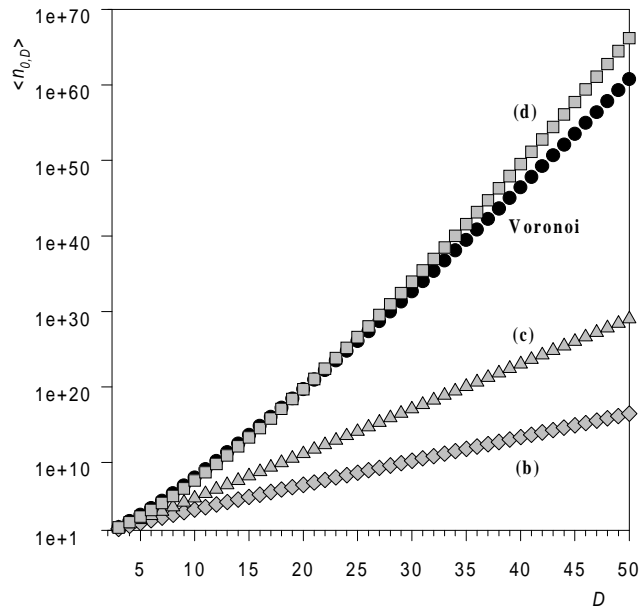}
\begin{centering}
\vspace{-3cm}
\caption{ \label{f.voronoi} The average number of vertices in a $D$ dimensional Vorono\"{\i} cell generated from Poissonian points can be exactly calculated [32] (black circles). Here this number is compared with the one associated with the fixed point structures (gray symbols). }
\end{centering}
\end{figure}

\end{document}